# INVESTIGATION OF STOCHASTIC RESONANCE (SR) NEAR HOMOCLINIC BIFURCATION IN A UNIJUNCTION TRANSISTOR (UJT) RELAXATION OSCILLATOR.

**Souvik Bose**\*#, **M.S. Janaki**\*\*, **Sandip Sarkar**\*\*\*

**A.N.Sekar Iyengar**\*\*

*\*Dept. of Astronomical Instrumentation, Indian Institute of Astrophysics ,Bangalore, India*

*\*\*Plasma Physics Division, Saha Institute of Nuclear Physics, Kolkata, India*

*\*\*\*Microelectronics Division, Saha Institute of Nuclear Physics, Kolkata, India*

#souvikbose@iiap.res.in

## Abstract

A P-spice simulation followed by an experiment with a unijunction transistor (UJT) has been carried out to investigate stochastic resonance (SR) in which the response of a nonlinear system to a weak periodic input signal is amplified by an optimum level of noise. The experiments were carried out in the vicinity of homoclinic bifurcation and the quantification of SR has been done by normal variance (NV) and signal to noise ratio (SNR) techniques. We have also developed a tentative mathematical model based on the current-voltage characteristic of the UJT and obtained a second order differential equation that was solved using MATLAB to yield a response similar to the one observed experimentally.

**Keywords:** Stochastic Resonance (SR), Unijunction Transistor (UJT), Signal to Noise Ratio (SNR), Differential Equation, Normalized Variance (NV)

# 1. <u>INTRODUCTION</u>

The phenomenon of detecting a sub threshold signal applied to a non linear system in the presence of noise (called SR) has been invoked in the context of a UJT based relaxation oscillator to explain the constructive effect of noise in excitable systems like the UJT. Furthermore there exists an optimum level of noise for which most effective detection takes place. Nurujjaman et.al.[1] observed the phenomenon of coherence resonance (CR) in a UJT, where it was found that there is a degradation of coherency of oscillations at higher noise amplitudes. This motivated us to advance and see if a UJT exhibits SR phenomenon as well.

Stochastic resonance has been studied in wide variety of physical, chemical as well as biological systems [1, 12- 16]. Parmananda et. al., [2] have investigated the effect of noise in electrochemical systems with excitable dynamics. It was found that there is a phenomenon of aperiodic stochastic resonance (ASR) in a chemical system exhibiting excitable dynamics. Effect of noise in excitable nonlinear systems like plasma has also been investigated by Nurujjaman et.al. [3]. In many of these systems, the source of non-linearity arises from the nature of the current (I)-voltage (V) characteristics. The UJT has a non-linear current voltage characteristics which is also the reason that they are used in triggering circuits in power electronics devices [5,6]. The study of noise induced resonance in UJT will hopefully help us to better understand the dynamics of a bistable electronic system like the UJT. They have a very interesting property that shows both limit cycle and fixed point behavior i.e. to say for a certain range of control parameter, limit cycle oscillations can be observed, while fixed point behavior is also observed for another range of values of the control parameter. Limit cycle is basically an isolated closed trajectory in phase space or an oscillatory behavior [4] in the time series. Since a UJT exhibits such a bistable behavior, it is interesting to find out if a weak undetectable signal can be detected once the UJT exhibits fixed point behavior. One of the ways to achieve this is by applying a weak periodic signal and random white Gaussian noise to the system. This phenomenon is called SR is quantified by estimating the Normal Variance (NV) which is the ratio between the standard deviation (std) ($t_p$) and mean ($t_p$), where $t_p$ is the interpeak time difference for the oscillatory time series. The NV helps us to measure the extent of regularity or coherency in the oscillations. This has been verified by signal to noise ratio method also.

It has been observed that there is an optimum level of noise when the response of the system yields most regular oscillations [1,12]. This fact is termed as coherency of the oscillations. Once the noise level reaches above a threshold value, degradation of coherency of oscillations is observed that might be explained as direct interference of the coherent oscillations with the input signals (noise & periodic signal) [7]. In the present work, SR is observed in a UJT by a P-SPICE simulation and also confirmed by experimental analysis. Finally, a tentative theoretical model has been developed based on the nonlinear current (I)-voltage (V) characteristic of the UJT, where homoclinic bifurcations [3] have been observed to develop from the relaxation oscillations, followed by the emergence of SR at the emitter terminal and it is quantified by the estimation of NV.

## 2. P-SPICE SIMULATION

A P-spice simulation of the basic relaxation oscillator circuit was carried out by using a Unijunction Transistor-Relaxation Oscillator (UJT-RO), which exhibits relaxation oscillations over a certain range of emitter voltages.

The simulation has been carried out with the circuit as shown in Fig 1. For the present case, the oscillator was operated at $V_{BB} = 5.0$ V. Depending on the voltage ($V_E$) between emitter (E) and base 1 (B1), periodic oscillations were observed and the output was recorded at emitter ($V_EO$). $V_E$, which is the control parameter in the present experiment can be varied by using a variable resistance $R_r$. Further we have coupled a noise source along with a weak periodic signal (a square wave) via the emitter terminal using two coupling capacitors for observing SR.

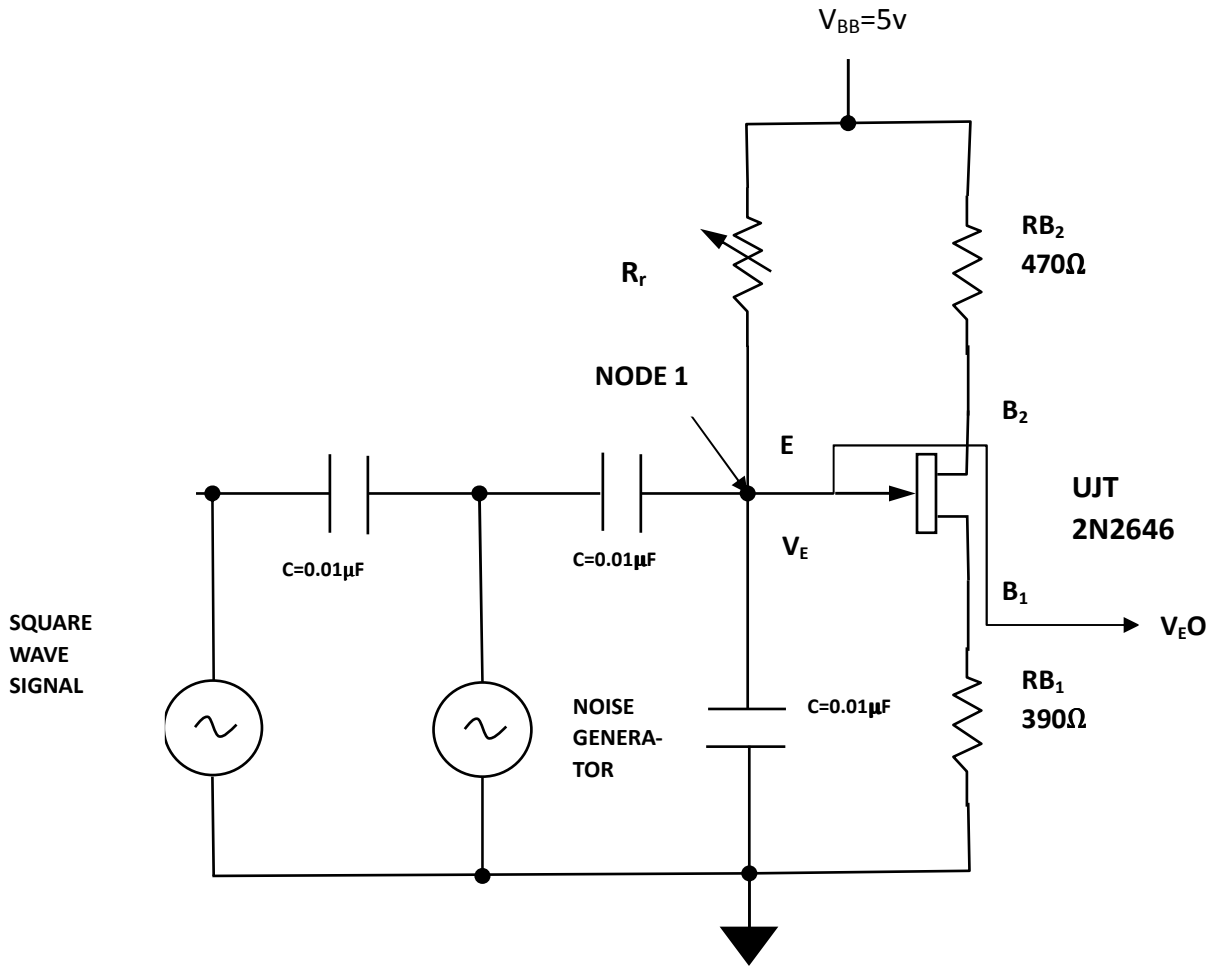

**Fig1. Circuit diagram for observing stochastic resonance using UJT-RO**

The relaxation oscillations are observed for a certain range of emitter voltage ($V_E$) which is controlled by the variable resistance ($R_r$) as shown in Fig 1. The relaxation oscillations were obtained between $V_E \approx 3.2V - 3.8V$. The time series plots of the relaxation oscillations obtained by using Pspice is shown in Fig 2 (a) -2 (c) below where it is clear that as the emitter voltage is increased the system reaches fixed point. It will be shown later, that the time period of the oscillations change, with the change in control parameter ($V_E$), and finally again after a certain threshold value, the UJT shows fixed point behavior.

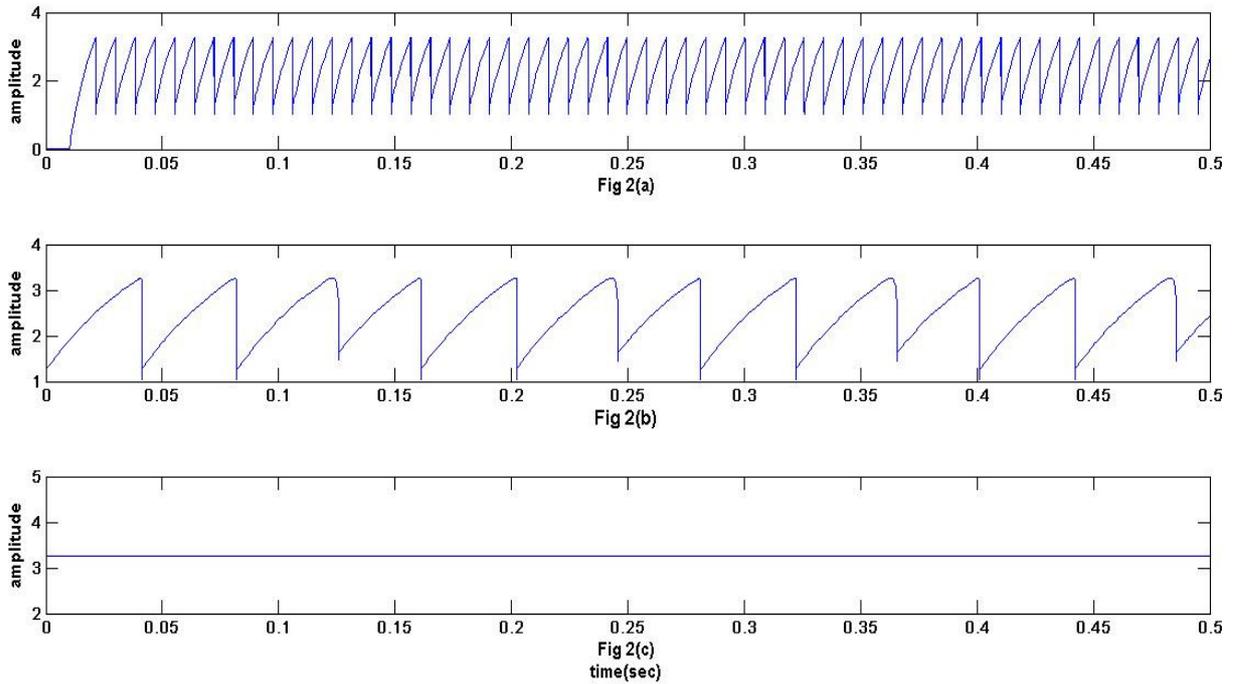

Fig 2 . Pspice bifurcation diagram constructed by increasing the control parameter $V_E$=3.2V (a), $V_E$=3.65V (b) and $V_E$=3.79 V (c) at the output terminal ($V_EO$).

To summarize, at values of $V_E$ below 3.2 V and above 3.8 V, excitable fixed point response was found. For our present experiments involving noise and a periodic signal the threshold was kept well above 3.8V and a weak nearly untraceable signal (square wave) was applied to the emitter terminal. The response of the system was noted after the inclusion of white Gaussian noise which was fed into the circuit using MATLAB.

For the experiments on Stochastic resonance (SR), the resistance $R_r$ was kept in such a way that the UJT exhibits fixed point behavior in the absence of noise and signal and a set point was chosen far from the threshold so that the system always remains in a stable state under the influence of intrinsic noise and parametric drifts [2]. To begin with, first a weak signal (here square wave) is superimposed by a coupling capacitor C as shown in Fig 1 at the emitter (E). The magnitude of the imposed signal should be such that it should not trigger the system which satisfies the criterion of being untraceable. So in order to satisfy this criterion, we found in our experiment that the square wave signal should not be more than 0.2mV peak to peak at a frequency of 0.002 KHz. Now at this stage a Gaussian noise generated by using MATLAB was applied via a second coupling capacitor of the same value as the initial one and the regularity of the provoked dynamics which now depends on the noise intensity were analyzed.

An important parameter used to quantify the observed behavior is the **Normalized Variance** (NV), (as mentioned before) and actually depicts the induced coherency. It is defined as NV= std($t_p$) / mean ($t_p$), where $t_p$ is the time interval between successive peaks in the time series oscillations. It is used to quantify the extent of induced regularity. It is evident that more coherent the induced oscillations, the lower the value of NV. For pure oscillations NV tends to zero.

Figures 3(b)-3(d) shows the time series of the output, taken from the emitter terminal ($V_EO$) of the UJT for different noise levels and Figure 3(e) represents the Normalized Variance (NV) as a function of noise amplitude in volts. We have the signal applied to the emitter terminal as shown in Fig 3(a). Point (1) in Fig 3(e) corresponds to the time series shown in Fig 3(b) and is associated with the low level noise where the activation threshold is seldom crossed, generating a sparsely populated irregular spike sequence. It is interesting to note that the spikes occur only during either the rising edge or the falling edge of the input square wave signal. Now as the noise amplitude increases, the NV falls as we can see from Fig 3(e) reaches a minimum as indicated by point (2) which corresponds to Fig 3(b) of the time series output from the emitter terminal. Here the response of the system is optimum i.e. a maximum regularity or uniformity is obtained in the time series output. There is a maximum level of coherency in this case, resulting in the fall of NV curve as was mentioned earlier. This is the region where the presence of a weak signal is detected. At higher amplitudes of the superimposed noise (input periodic signal being constant in amplitude), the observed regularity in the time series is destroyed as manifested by an increase in NV; labeled as (3) in Fig 3(e), and the corresponding time series has been shown in figure 3(d). This is a consequence of the dynamics being dominated by the noise. So we can justify the definition of stochastic resonance which means that at a finite optimum level of noise the response of the system is maximum **[8]**, which will be later on verified by a plot of SNR (Signal to noise ratio) as a function of noise amplitude.

Fig 3(e) shows that the system attains a state where coherency of oscillations increases with an increase in amplitude of the noise initially and stays at that state for a wide range of amplitudes before being degraded at higher values.

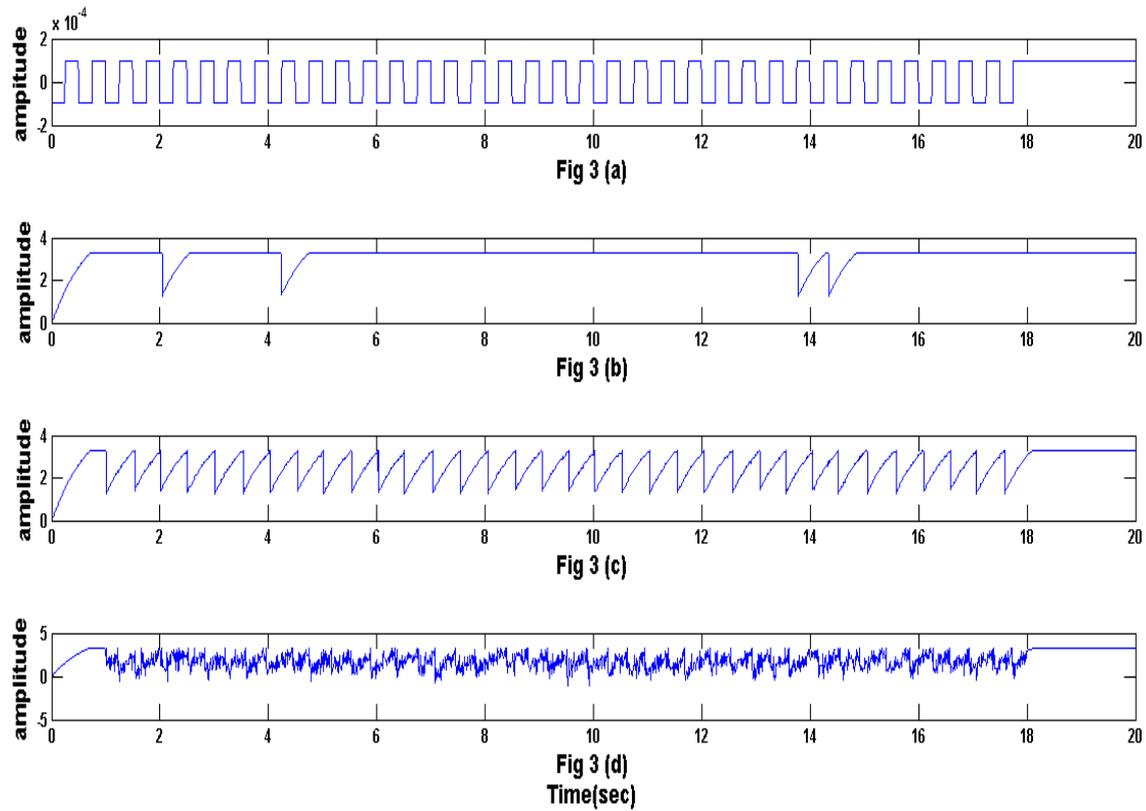

**Fig 3(a)-(d) showing the emergence of stochastic resonance recorded at emitter terminal ($V_E$) for (b) low (0.001 V), (c) optimum (0.058 V) and (d) high (2 V) level noise.**

Initially the inter peak distance or time is dominated by the activation time of the limit cycle oscillations, which varies significantly at low noise levels, and with an increase in the noise amplitudes the activation time decreases resulting in a rapid fall of NV [**9,10**]. When the noise amplitude is such that it crosses threshold very often, i.e. number of crossings is much larger than the excursion frequency of the oscillations, it remains in the excited state, resulting in coherent oscillations leading to a flat minimum in the NV curve as shown in Fig 3(e) below [**2**]. This region is characterized by the presence of the weak periodic signal. Direct interference of strong noise distorts the limit cycle oscillations, particularly the peaks, and since the excursion time is almost independent of the noise amplitude [**9,10**] the increase in NV at higher noise amplitudes is probably due to the increase in the variation of the inter peak distances of the distorted peaks of the limit cycle oscillations [**7**].

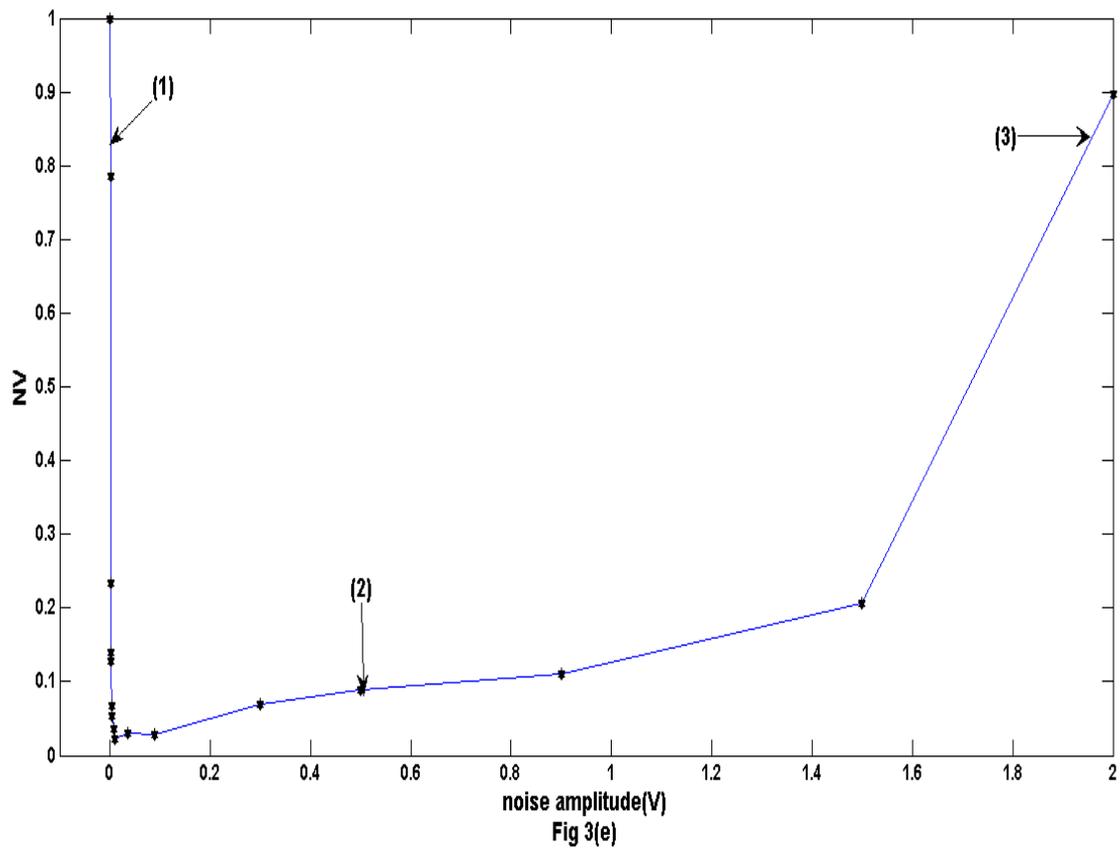

**Fig 3(e) shows NV as a function of noise amplitude (V) for the experiments performed at $V_E$=4 V**

## 3. **EXPERIMENTAL ANALYSIS**

First of all we would like to focus on the relaxation oscillations of the simple UJT-RO obtained only within a certain range of emitter voltage ($V_E$) **[3]**. Here in the actual experiment we get the oscillations from $V_E$ = 3.05V to $V_E$ = 3.69V controlled by using the variable resistance $R_r$ .We explain the phenomenon here by the construction of a bifurcation diagram as shown below.

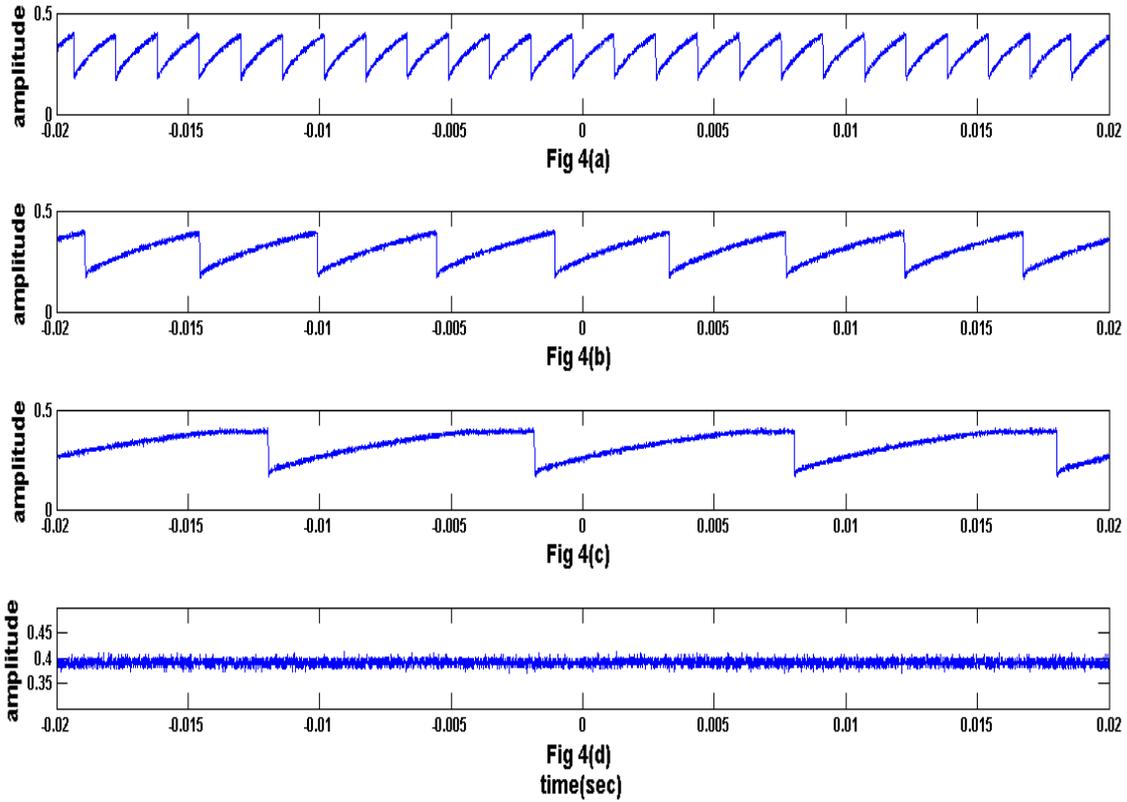

**Fig 4. Experimental bifurcation diagram constructed by increasing the control parameter $V_E$=3.05 V (a), $V_E$= 3.28 V (b), $V_E$= 3.45 V (c) and $V_E$=3.69 V (d) at the output terminal ($V_EO$).**

It can be seen from above diagram that on increasing the voltage at the emitter the oscillations die out with a gradual increase in time period (T) signifying homoclinic bifurcation. The variation of (T) with $R_r$ is given by the standard equation of time period of UJT as follows [5]

$$T = R_r C \ln(\frac{1}{1-\eta}) \qquad (1)$$

where η is called the intrinsic stand-off ratio lying between 0-1. The time period also increases with $V_E$. As found out by Nurrujaman *et al*.[1] the relationship between lnT and ln$V_E$, follows a power law behavior with an exponent of ≈ 0.99. So as we vary the control parameter $V_E$, relaxation oscillations were observed in the range of ≈ 3.05 V to 3.69 V. So $V_E$ ≈ 3.7 V defines the threshold of the system.

Here, again we set the threshold of the system much higher than 3.7 V, for noise and signal invoked dynamics exactly the same way we did for our P-SPICE simulation. For our present case we have found out that the sub threshold periodic signal ( a square wave ) should be of 8 mV peak to peak in amplitude at a frequency of 0.032KHz which is to be coupled to the emitter by a coupling capacitor (C= 0.01 µF) using a function generator. At this stage we start applying white Gaussian Noise by another function generator via a second coupling capacitor in exactly the same way we had done for the simulation.

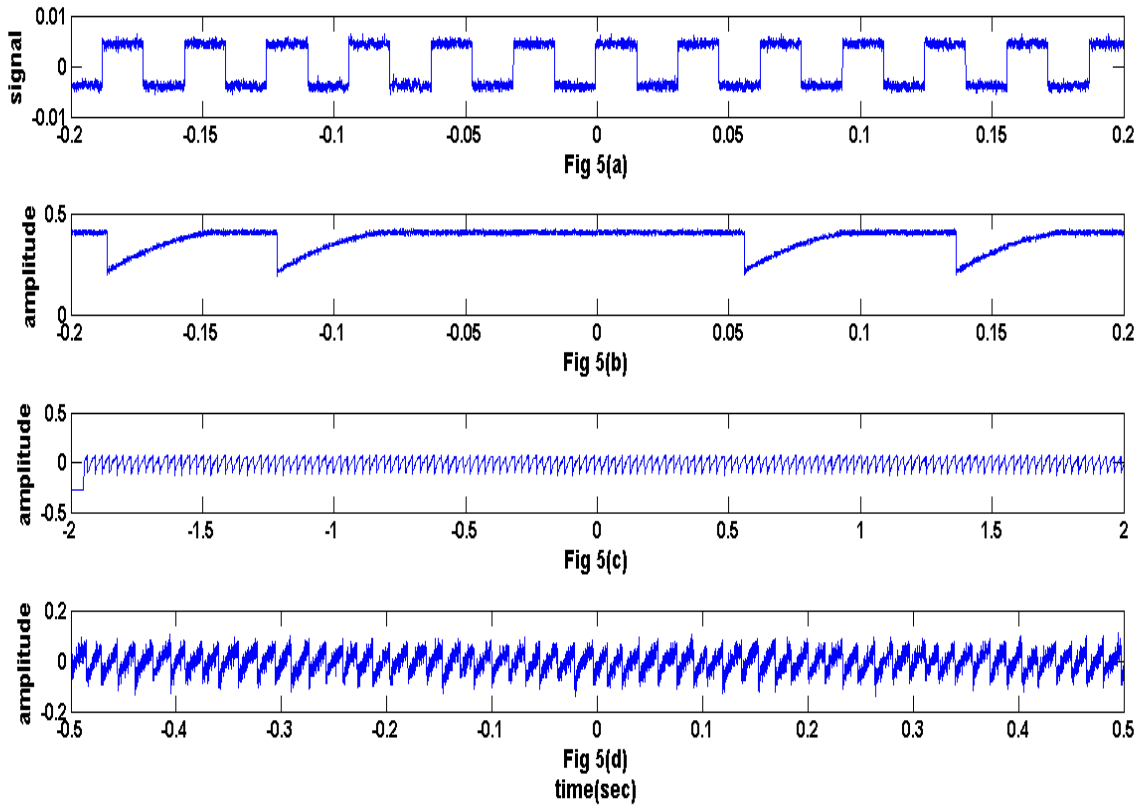

**Fig 5(a)-5(d) showing the emergence of stochastic resonance recorded at emitter terminal ($V_E$) for (b) low (0.08 V), (c) optimum (1.5 V) and (d) high (6 V) level noise.**

Again as in our previous analysis for the Spice simulation, we have quantified the above experimental bifurcation diagram by a NV vs. noise amplitude plot. The explanations are exactly similar to the ones already discussed in the previous sections.

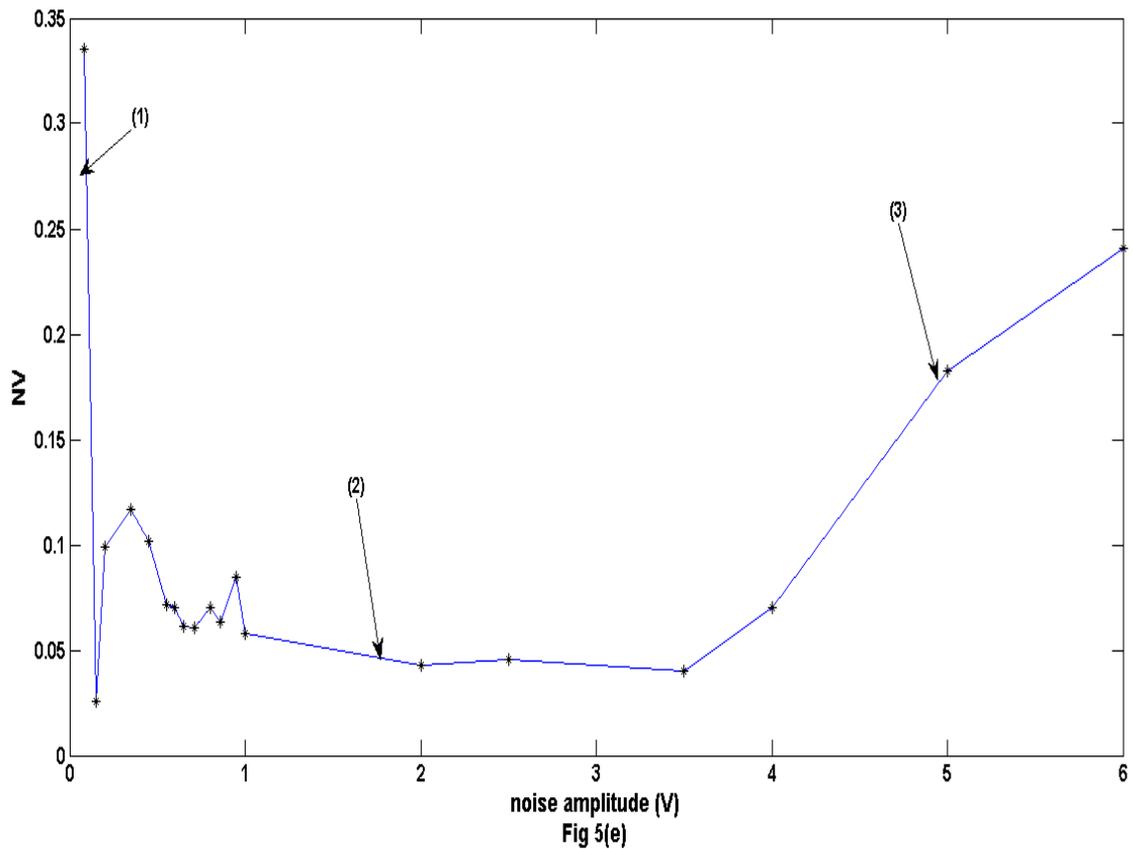

**Fig 5(e) shows NV as a function of noise amplitude (V) for the experiments performed at $V_E$=3.75 V**

Points (1), (2) & (3) in Fig 5(e) corresponds to oscillations in Fig 5(b), 5(c) & 5(d) respectively. Here, as expected we find that that the NV initially falls rapidly which implies that the coherency of oscillations increases until it is degraded by the direct interference of noise at a very high value of noise voltage [**7**].

Our experimental results follow our simulation results closely. The NV curves and the time series outputs are similar for both the experiment and the simulation.

## 4. <u>Signal-to-Noise ratio (SNR) in stochastic resonance using UJT.</u>

The concept of Signal to Noise ratio (SNR) is quite significant in qualitative analysis of stochastic resonance phenomenon observed in our experiment. SNR theoretically means the ratio between average signal power to average noise power. According to the definition of stochastic resonance, there is a finite optimum level of noise at which the response of the system is the "best". Here the frequencies of the weak periodic square signal match the random frequency of the noise. Therefore we obtain a peak in the response of the SNR. So we can say that at a certain instant, the signal strength is maximum which suppresses the input noise [**8**]. As can be seen from the figures above i.e. Fig 5(b)-5(d), the oscillatory response is the best out of the three for an intermediate level (1.5 V) of noise (5(c)) and decays both before and after a certain level (Fig 5(b) & 5(d)). Therefore we can obtain a figure of SNR (dB) vs. noise amplitude for our experimental data in the following form.

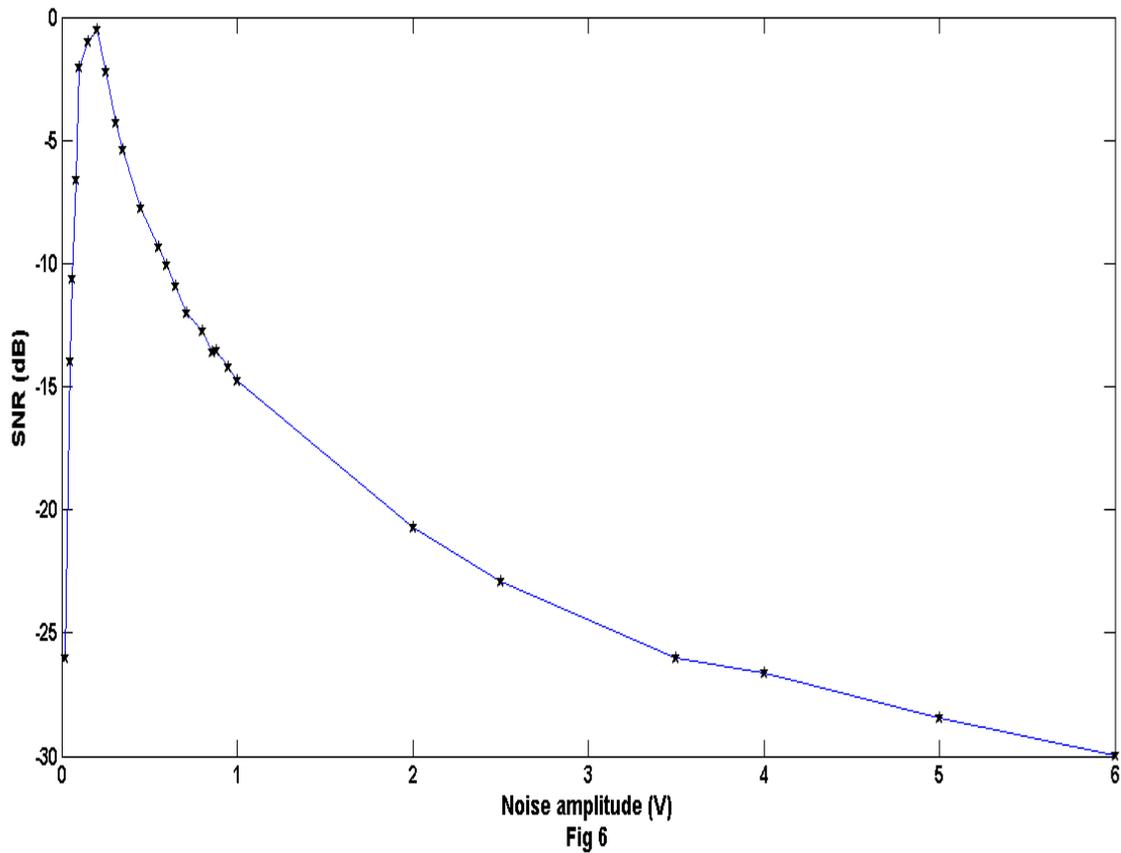

Fig 6

**Fig 6 shows SNR (dB) as a function of Noise amplitude (V).**

Fig. 6 indicates clearly that there is a "peak" response of the oscillations obtained from the emitter terminal for a certain value of the noise amplitude. This quantifies the phenomenon of stochastic resonance.[**11**]

# 5. Proposed Theoretical Model for the UJT based Relaxation oscillator on the basis of its I-V characteristics.

The V-I characteristic of a UJT can be divided into three regions [**5**] - 1.The cut-off region. 2. Negative resistance region 3.Saturation region. In Fig 7(a) the practically obtained V-I characteristic is shown along with the polynomial fitting. In the cut-off region a very small amount of current (in nano ampere range) flows through the E-$B_2$ junction (Fig 1) due to its reverse bias condition. To counteract this reverse bias the emitter potential ($V_E$) increases unless it reaches the **peak voltage** $V_P$ = $V_D$+ η$V_{BB}$, where $V_D$ is around 0.5-0.7 V, η is the intrinsic standoff ratio and the $V_{BB}$ is the bias voltage. Till then the capacitor (C) charges through the variable resistance ($R_r$) path. After $V_E$ (emitter potential) crosses $V_P$, there is a sudden decrease in the emitter voltage followed by an increase in the emitter current due to a rapid fall of the internal base resistance in $B_1$. This is the onset of the negative resistance region of the UJT i.e. it is said to be triggered. At this stage, the capacitor discharges though the UJT via E-$B_1$ junction (Fig 1) and the emitter potential decreases. This decrease in $V_E$ continues till it reaches the **valley point** ($V_V$) or the point with the minimum value of emitter potential. After this the UJT is cut-off again which again results in charging of the capacitor (C). This process of switching between cut-off mode and negative resistance mode continues till the **saturation** is reached which is characterized by an increase in emitter current along with an increase in emitter potential. The valley point is the minimum point of the characteristic curve. It is easy to understand that the characteristic is non-linear which indicates that the UJT is a non-linear system.

We plot the co-ordinates of the voltage (V) and current (I) obtained from our experiment. And fit it to obtain a mathematical expression of the same. For simplicity, we have assumed a polynomial fit.

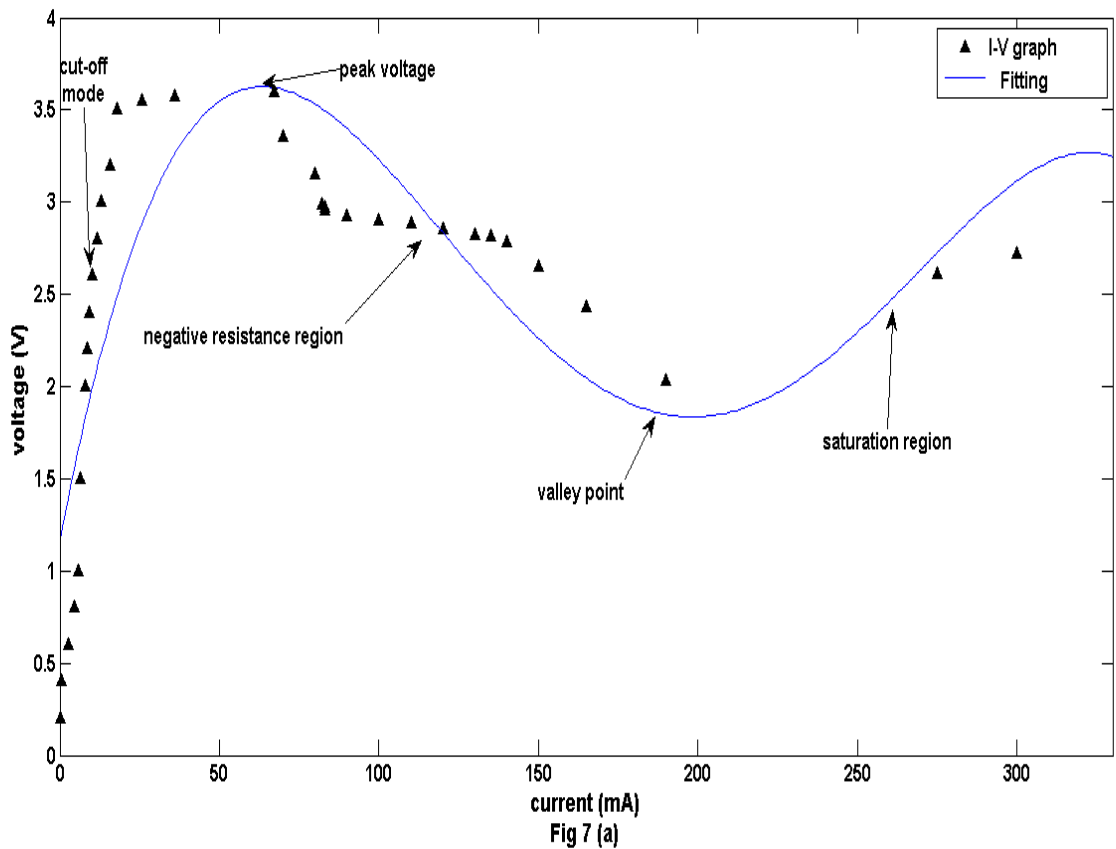

**Fig. 7 (a): Emitter voltage versus current obtained experimentally and its approximate polynomial fitting.**

The above fitting curve is expressed in the form of a polynomial equation of voltage as a function of current in the form as- $V = aI^4 + bI^3 + cI^2 + dI + e$. We have considered a 4$^{th}$ order fitting since it is the closest approximation to the experimental characteristics as shown below in Fig 7(b).

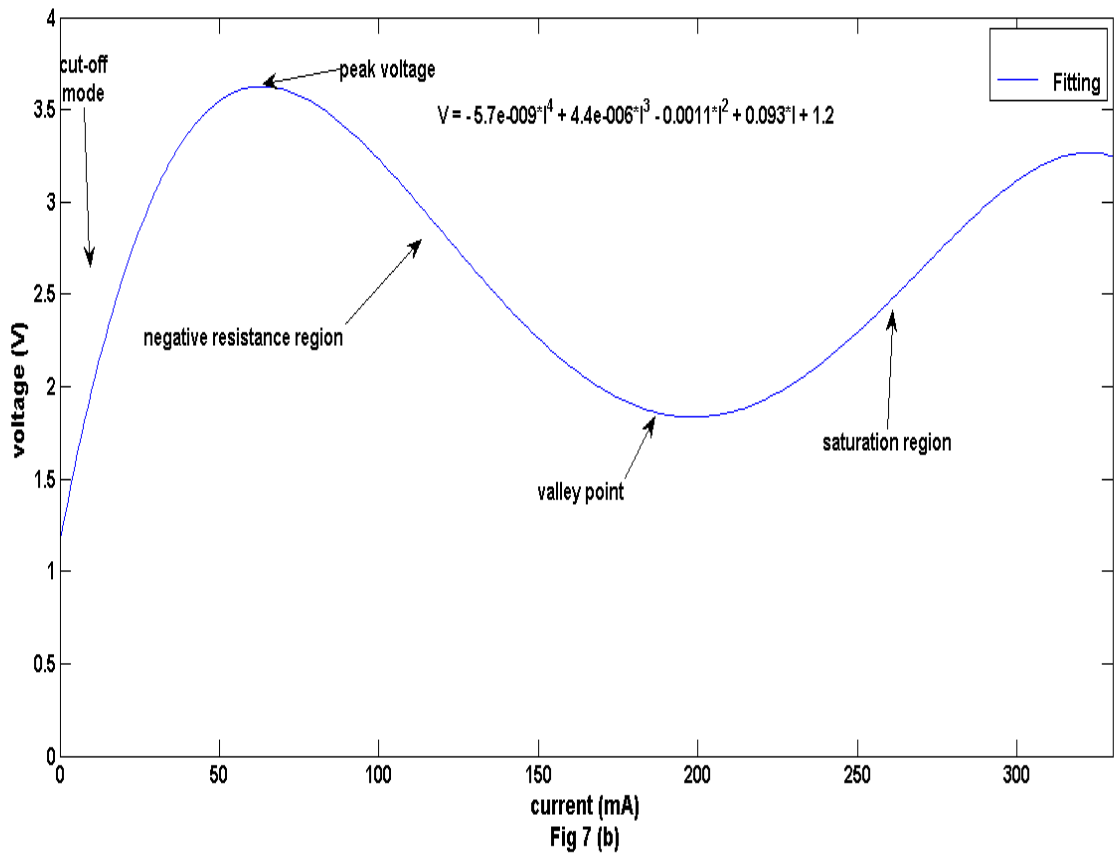

**Fig 7(b): V vs. I polynomial expression.**

Our basic motive behind the fitting was to find **current** in terms of **voltage** (instead of voltage in terms of current) to aid our further calculations. We have tried to derive an approximate curve of the current (I) in terms of Voltage (V) and expressed it also as a fourth order polynomial to simplify the analysis. Our analysis is shown separately in Fig8 below. This may not be the typical inverse of Fig 7(b), however it does serve our purpose of mathematical analysis to a very high degree. The actual I vs. V curve, is shown in blue color whereas the fitted polynomial curve is shown in red. It is to be noted that like previous analysis we have restricted ourselves to polynomial fitting of the curve to simplify the derivations that are followed.

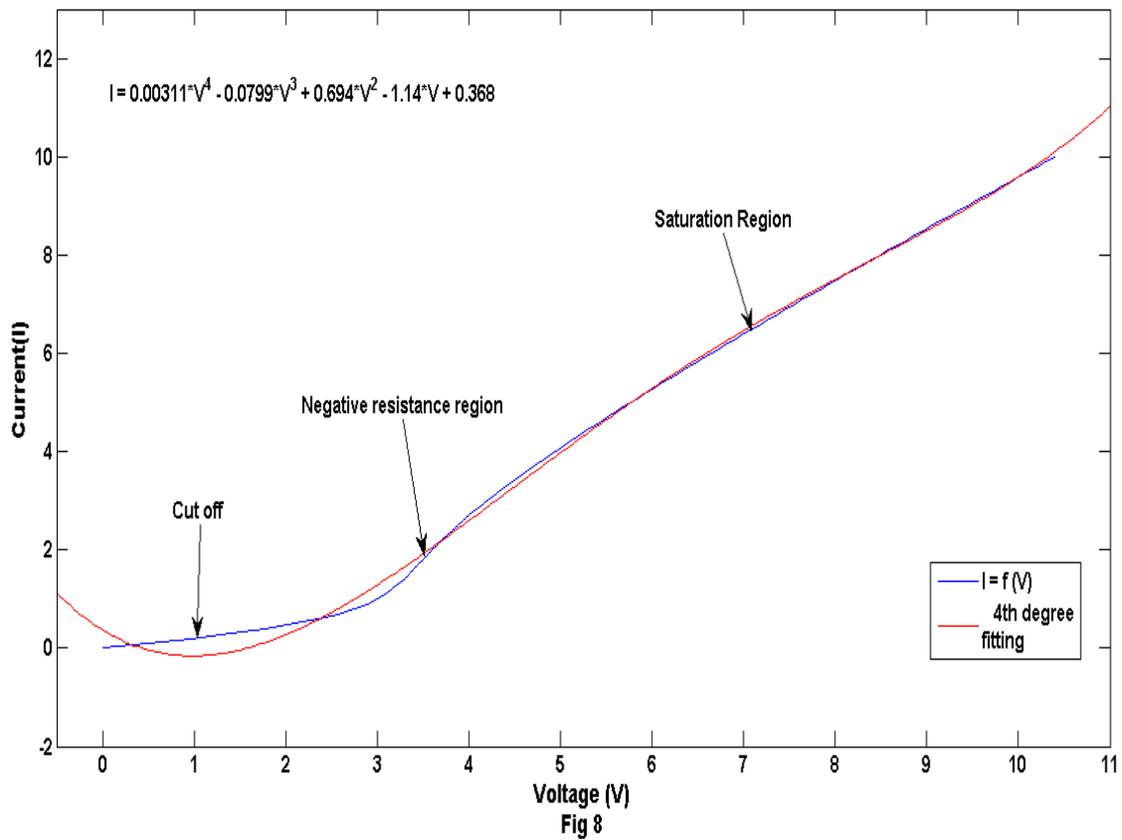

**Fig 8: I vs. V polynomial fitting.**

The polynomial is of the form of **ax⁴+bx³+cx²+dx+e** where a, b, c, d & e have the values as shown above in the figure. Therefore current can be expressed in terms of the voltage in the following way

$I = \varphi(V) = aV^4 + bV^3 + cV^2 + dV + e;$  (2.1)

For circuit analysis, we would like to redraw the circuit of Fig 1 in an alternative way with an intention of applying the basic laws of Circuit theory for simplifying the analysis. Therefore the UJT-RO circuit can be redrawn as shown in Fig 9. Here, we stress on the fact that we have treated the UJT as a whole in the rectangular box. Our main intention is to analyze the complete relaxation oscillator circuit as a whole. The following figure is a simplified version of the original circuit of Fig 1.

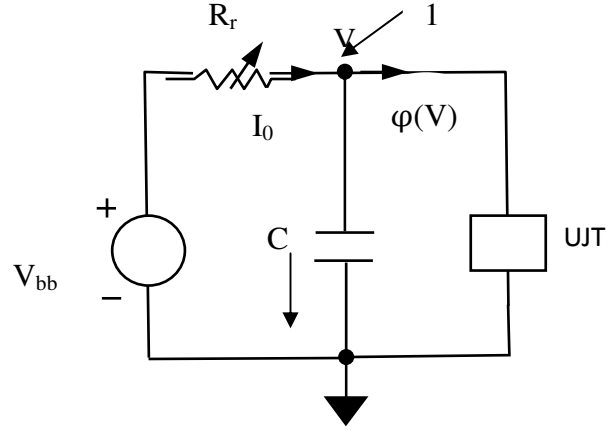

**Fig 9 - Equivalent circuit diagram for the UJT-RO relaxation oscillator**

From the circuit shown in Fig 9, we want to apply Kirchhoff's current law at node 1 (having a potential say V). Let $I_0$ be the current flowing into node 1 from the bias voltage $V_{bb}$,

We have

$$I_0 = C\frac{dV}{dt} + \varphi(V) \qquad (2.2)$$

$$\therefore C\frac{dV}{dt} = I_0 - \varphi(V) \qquad (2.3)$$

Now differentiating equation (2.3) again with respect to time t, we have

$$C\frac{d^2V}{dt^2} = \frac{dI_0}{dt} - \frac{d\varphi(V)}{dt} \qquad (2.4)$$

It is to be noted that we have performed the second derivative to find out the variation of V with respect to time in $\varphi(V)$.

$$\frac{d\phi(V)}{dt} = \frac{d\phi(V)}{dV} \times \frac{dV}{dt}$$

φ (V) can be substituted by equation (2.1) above and differentiated to obtain the following

$$\therefore \frac{d\varphi(V)}{dt} = \frac{dV}{dt}(4aV^3 + 3bV^2 + 2cV + d) \qquad (2.5)$$

An expression for $\frac{dI_0}{dt}$ can be obtained, from the nodal analysis at (node) 1,

Such that the magnitude of $I_0 = (V - V_{bb})/R_r$

Here V and $R_r$ both are variable with respect to time because by varying $R_r$ (control parameter) the value of potential at node 1 (V) is controlled. Also, since we considered the UJT as a box in the above analysis, we have modeled $R_r$ in such a way that it will depend on the internal resistance of the UJT which is varying with time as follows-

$$R_r = R_{ext} + r_{internal-ujt}(t)$$

Here we have assumed that $R_{ext} \gg r_{internal-ujt}(t)$ generally. So while including $R_r$ in the numerical analysis using MATLAB later on, we can write $R_r \approx R_{ext}$ and neglect $r_{internal-ujt}(t)$ which is assumed to be very small with respect to $R_{ext}$. However, while differentiating $R_r$ as given by the above equation, we write

$$\frac{dR_r}{dt} = \frac{dr_{internal-ujt}(t)}{dt} \approx k \quad [\text{k is assumed to be a constant}]$$

Now for our analysis it is assumed that k=1, which after substituting in $\frac{dI_0}{dt}$ we get the following equation-

$$\therefore \frac{dI_0}{dt} = \frac{dV}{dt}\frac{1}{R_r} - \frac{(V-V_{bb})}{R_r^2} \tag{2.6}$$

Finally by substituting **(2.5)** & **(2.6)** in equation **(2.4)**,

We obtain the oscillation/voltage equation as

$$C\frac{d^2V}{dt^2} = -\frac{dV}{dt}(4aV^3 + 3bV^2 + 2cV + d - \frac{1}{R_r}) + \frac{(V_{bb}-V)}{R_r^2} \tag{2.7}$$

The coefficients have their values as given in the fitting curve in Fig. 8 above.

Now the above second order differential equation is solved using MATLAB with slight adjustment in the coefficients and by adjusting and varying the parameters like $R_r$ (resistance) & C (capacitance) [the control parameters] we get the relaxation oscillations first, then for a change in the parameters, we can observe the homoclinic bifurcations, where the time period increases and finally fixed point behavior is observed exactly in a similar fashion as in the cases of Spice simulation and experimental analysis. These can be shown in the following set of figures for different value of $R_r$. So we proceed in the same way as we did before. First, we show the theoretically obtained bifurcation diagram from the above equation followed by the emergence of stochastic resonance at the emitter terminal by the application of a sub threshold periodic square wave and noise.

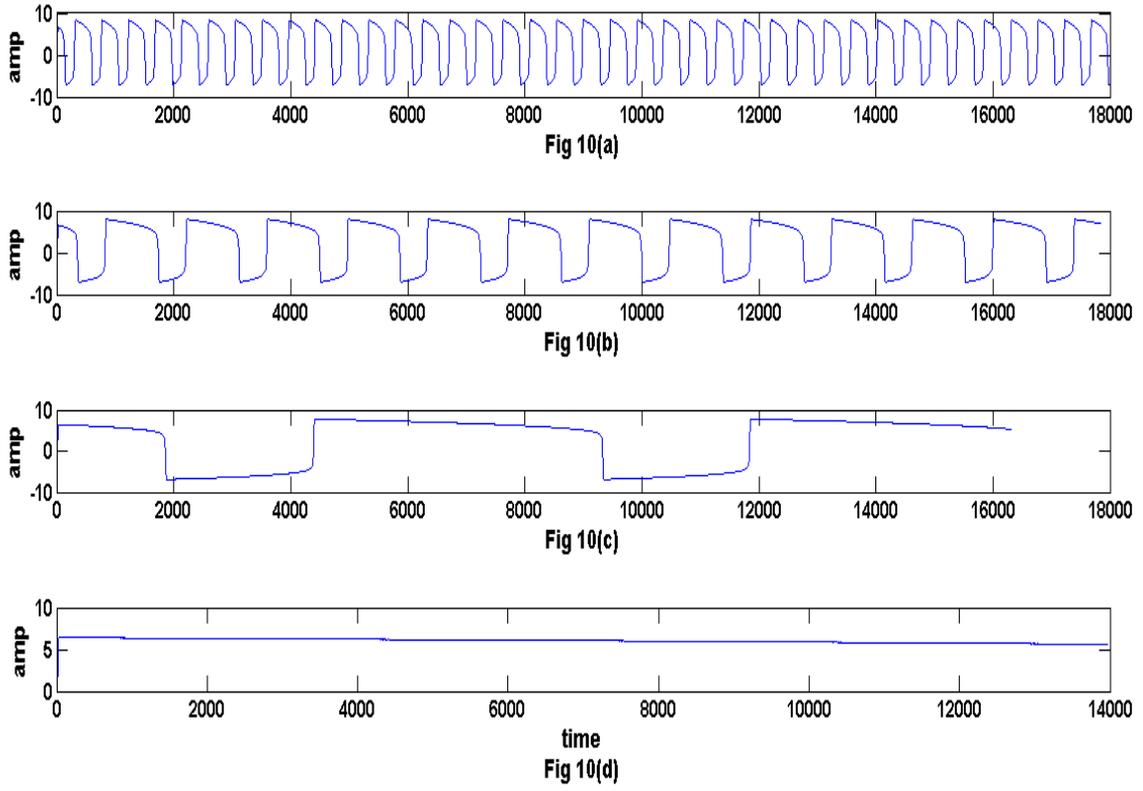

**Fig 10. Theoretical bifurcation diagram obtained by varying the control parameter $R_r$ =10 Ohms (a), $R_r$=35 Ohms (b), $R_r$= 75 Ohms (c) and $R_r$= 92 Ohms (d)**

Now at the fixed point, we introduce noise and very weak square waves (periodic signal) fed through MATLAB and try to observe stochastic resonance. Theoretically, that would mean adding a noise term and a signal term to the right hand side of equation **(2.7)**. Therefore **(2.7)** can be written as follows

$$C\frac{d^2V}{dt^2}+\frac{dV}{dt}(4aV^3+3bV^2+2cV+d-\frac{1}{R_r})-\frac{(V_{bb}-V)}{R_r^2}=0$$

Now adding noise term **D ξ(t)** and signal term **A ψ(t)** to the R.H.S. of the above equation we have

$$C\frac{d^2V}{dt^2}+\frac{dV}{dt}(4aV^3+3bV^2+2cV+d-\frac{1}{R_r})-\frac{(V_{bb}-V)}{R_r^2}=\mathbf{A}\,\psi(\mathbf{t})+\mathbf{D}\xi(\mathbf{t}) \qquad (2.8)$$

Where A and D are the amplitudes of periodic signal and noise respectively.

Gradually we increase the amplitude of noise (D) and find out the emergence of stochastic resonance at the emitter terminal as follows

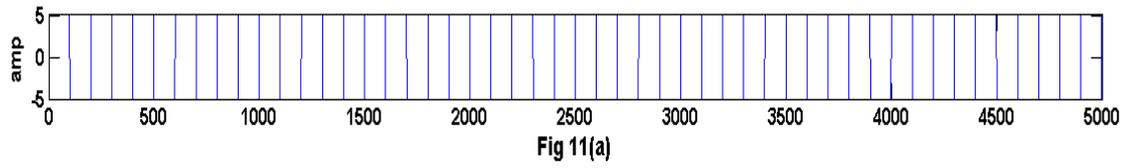
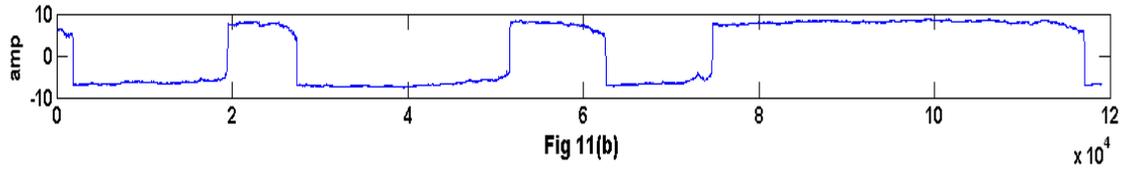
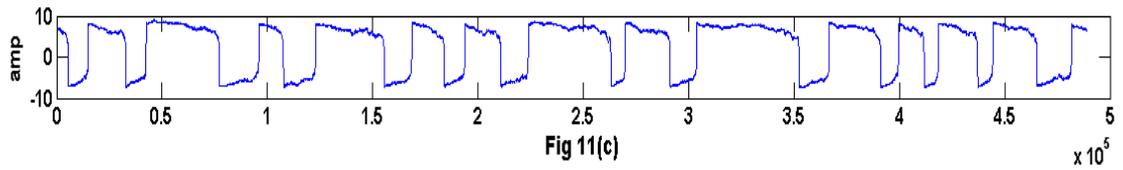
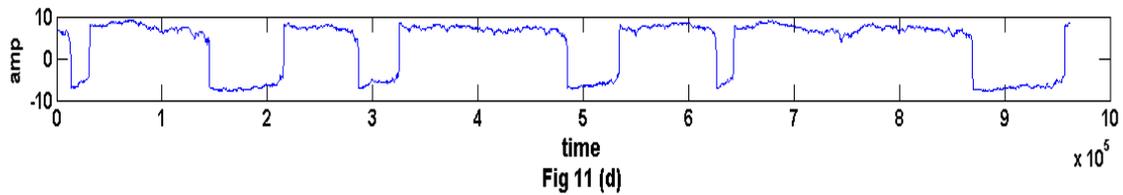

**Fig 11(a)-11(d) showing the emergence of stochastic resonance recorded at emitter terminal ($V_E$) for (b) low (0.006 V), (c) optimum (0.0028 V) and (d) high (0.01 V) level noise.**

As in the case of the Pspice simulation and experimental analysis, we would also like to quantify our observation of stochastic resonance by plotting the corresponding NV vs. Noise amplitude plot for our theoretical prediction. The plot and its correspondence with the above observation is presented in Fig 11(e) below

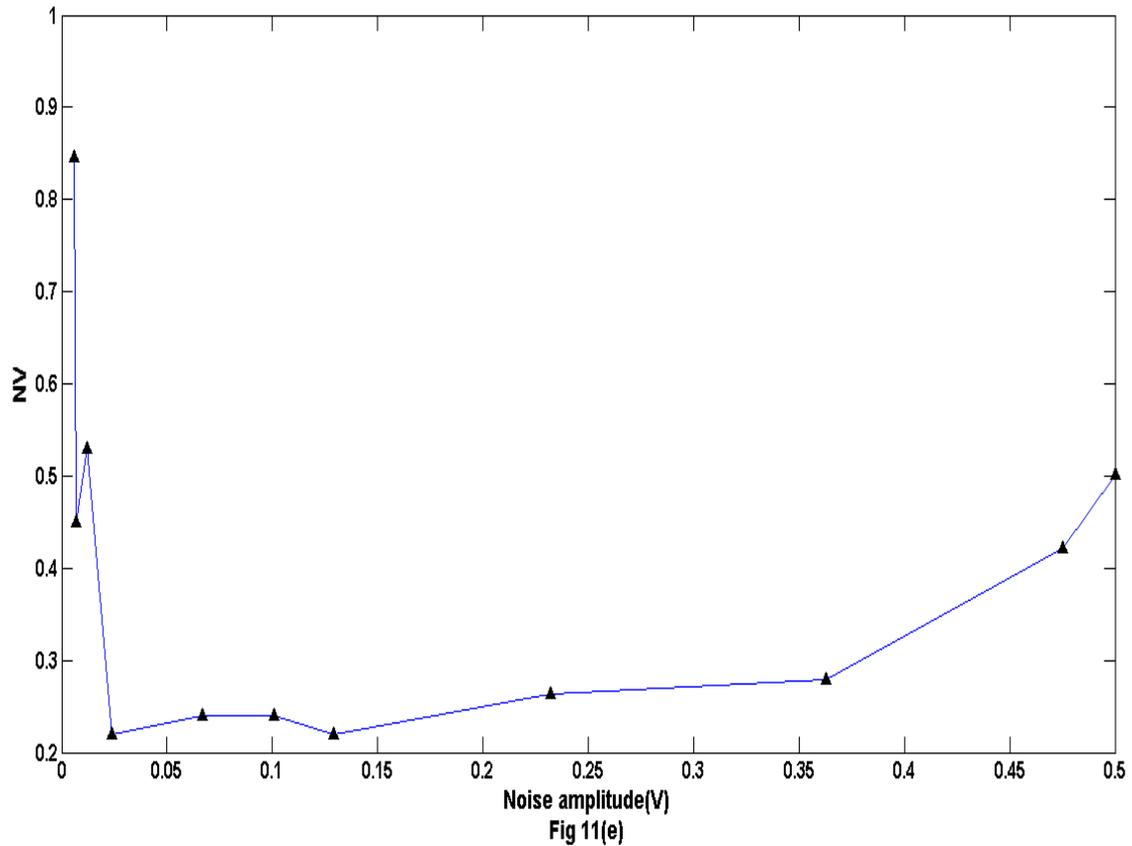

**Fig 11(e) shows theoretically obtained NV vs. Noise amplitude curve**.

The NV curve and the time series for the theoretical results show features almost similar to the experimental results. It is to be noted that the above theoretical model is only an attempt to understand the dynamics of the UJT based relaxation oscillator based solely on the current-voltage characteristics.

# 6. CONCLUSION

We have studied the effect of random forcing (noise) along with periodic forcing (square wave) in a bistable system like the UJT-RO, and demonstrated the phenomena of stochastic resonance both experimentally and by using P-SPICE simulations. Further, we developed an approximate theoretical model explaining the above phenomenon. Our results show almost similar behavior in all the schemes of analysis. We have also shown that Signal-to-noise ratio (SNR) vs. noise amplitude plot has a maximum for a certain optimum level of noise amplitude which should be exactly the case for a stochastic resonance phenomenon. This concept of stochastic resonance may well be applied in transmission of sub threshold signals which are generally very difficult to detect. We can apply an optimum level of noise & periodic forcing to the signal to maximize the response of the system. The results of the simulation have been very encouraging which compels us to believe that stochastic resonance can be observed for any device which is having a negative resistance region in its current-voltage (I-V) characteristic. This leaves room for applying this concept to various other devices like tunnel diodes and Gunn Diodes.

## 7. Acknowledgements

The authors would like to acknowledge the technical assistance help from the members of the Plasma Physics Division, Saha Institute of Nuclear Physics, Kolkata. The authors would like to thank Dr. Koushik Ghosh and Mr. Partha Sarathi Chakraborty of the University Institute of Technology, Burdwan for their valuable support and inspiration.

## References

[1] M. Nurujjaman *et al.,* Phys. Rev. E **80,** 015201(R) (2009).

[2] P. Parmananda, Gerardo J. Escalera Santos, M. Rivera, and Kenneth Showalter, Phys. Rev. E **71,** 031110 (2005)

[3] M. Nurujjaman, A.N. Sekar Iyengar & P. Parmananda, Phys. Rev. E. **78,** 026406 (2008)

[4] Strogatz S.H., *Nonlinear Dynamics & Chaos* ( *Levant Books, Kolkata, India, 2007)*

[5] M.Allen, *Electronic Devices and Circuits* (Prentice Hall Of India, Delhi, 1980), pp 538-540.

[6] M.I.M. Fetiosa and O.N.Mesquita, Phys. Rev. A. **44,** 6677 (1991)

[7] G. Giacomelli *et al.,* Phys. Rev. Lett. **84,** 3298 (2000)

[8] Roberto Benzi, Alfonso Sutera, and Angelo Vulpiani, J. Phys. A **14,** L453 (1981).

[9] A.S. Pikovsky and J. Kurths, Phys. Rev. Lett. **78,** 775 (1997)

[10] K. Miyakawa & H. Isikawa , Phys. Rev. E. **57,** 3292 (1998)

[11] Bruce McNamara and Kurt Wiesenfeld, Phys. Rev. A, **39,** 9, (1989)

[12] Bruce McNamara, Kurt Wiesenfeld, and Rajarshi Roy, Phys. Rev. E **60,** 2626 (1998).

[13] F.Moss, A.Bulsara, and M.F Shlesinger, J.Stat.Phys. **70,** 1 (1993)

[14] A.Longtin, A.Bulsara, and F.Moss, Phys. Rev. Lett. **67**, 656 (1991).

[15] Gerardo J. Escalera Santos and P.Parmananda, Phys. Rev. E **65,** 067203 (2002).

[16] T.Amemiya, T. Ohmori , M.Nakaiawa, and T. Yamaguchi J. Phys. Chem. **102**, 4537 (1998).

.